\documentclass[preprint,aps]{revtex4}
\usepackage{amsmath,amssymb}
\usepackage{graphicx}
\usepackage{dcolumn}
\usepackage{bm}
\usepackage{epstopdf}
\begin{document}
\author{S. F. Caballero Ben\'\i tez$^{1}$, V. Romero-Roch\'{\i}n$^2$ and R. Paredes$^2$,} 
\affiliation{
$^{1}$ARC Centre of Excellence for Quantum-Atom Optics and Nonlinear Physics Centre, Research School of Physical Sciences and Engineering, Australian National 
University, Canberra ACT 0200, Australia
\\
$^{2}$ 
 Instituto de F\'{\i}sica, Universidad
Nacional Aut\'onoma de M\'exico, Apdo. Postal 20-364, M\'exico D.
F. 01000, M\'exico. }

\title{ Delocalization to self-trapping transition of a Bose fluid confined in a double well potential. An analysis via one- and two-body correlation properties}
\date{\today}
\begin{abstract}
We revisit the coherent or delocalized to self-trapping transition in an interacting bosonic quantum  fluid  confined in a double well potential, in the context of full quantum calculations. We show that an $N$-particle Bose-Hubbard fluid reaches an stationary state through the two-body interactions. These stationary states are either delocalized or self-trapped in one of the wells, the former appearing as coherent oscillations in the mean-field approximation. By studying one- and two-body properties in the energy eigenstates and in a set of coherent states, we show that  the delocalized to self-trapped transition occurs as a function of the energy of the fluid, provided the interparticle interaction is above a critical or threshold value. We argue that this is a type of symmetry-breaking continuous phase transition.
\end{abstract}
\maketitle

\section{Introduction.} 

The experimental realization of  a single bosonic Josephson junction in two weakly linked Bose-Einstein condensates (BEC) has revealed fundamental features of the dynamical nature of an ultracold interacting Bose gas in a double-well potential\cite{Oberthaler}. The atomic transport has shown the existence of Josephson tunneling or macroscopic quantum self-trapping regimes depending on the interatomic interactions and on the initial populations in the wells. These opposite states are similar to the well known superfluid and Mott insulator states observed in optical lattices \cite{I.Bloch,Teichmann} in the sense that the quantum transition among them is the result of the non-linear condensate self-trapping interactions. 

The theoretical analysis of this system has been based on the Bose-Hubbard Hamiltonian in the two-mode approximation\cite{Milburn,Holthaus,meanfield1,meanfield2,Paredes1,Paredes2}. Besides providing a fairly good description of the experimental situation\cite{Oberthaler}, this model has been extensively studied on its own, specially within a mean-field or semiclassical approximation. This model has lead to significant understanding of the richness of the physical problem at hand; additionally, it can incorporate asymmetric two-well potentials and external time-dependent driving fields\cite{Holthaus}.

Beyond the studies within the mean-field approximation, this model is amenable to numerical full quantum calculations, so far up to $N = 1000$ particles\cite{Milburn,Holthaus,Paredes1,Paredes2}. These results and mean-field calculations agree for a short time while for longer times the full quantum approach shows that the system reaches a state, that one may call ``statistically stationary" as we shall specify below,  punctuated by coherent revivals. Here, we revisit the symmetric two-mode Bose-Hubbard model within numerically exact quantum calculations to, first, verify the previous results and, second, to accomplish a more detailed description of the self trapping transition.  

The mean-field approach, which approximates the $N$-body dynamics by a set of non-linear coupled dynamical equations, is limited and incomplete since it describes the evolution of the expectation values of one-body operators only.  However, from the full quantum dynamics one may, in principle, enquire about the dynamics of one-, two-, three-, and up to $N$-body properties. Due to the relative simplicity of the present model one can calculate either the full $N$-body wave function for a pure state or the full $N$-body density matrix for a mixed state (for a system up to  $N =$ 10,000 particles, say). Thus,  full information of the state of the system may be obtained.

Anyhow, from a macroscopic point of view, most of the measurable thermodynamic and/or many-body properties, such as energy, temperature and Green's functions in general, are given in terms of one- and two-body operators\cite{Fetter}. These in turn are exhaustively described by the one- and two-body reduced density matrices. Although we are able to calculate the complete one- and two-body density matrices from the exact solution, here for purposes of showing our main points we shall limit our study  to the following measurable quantities: the number of particles in one of the wells, given by the expectation value of
a one-body operator  $ \hat N_1$, and a tunneling correlation function, the expectation value of a two-body operator $\hat J_x \hat J_x$. Below we give the explicit form of these operators.

Our results may be summarized as follows. Since we are dealing with a macroscopic system, following the principles of statistical mechanics, we choose to describe the state of the system by specifying the number of particles $N$ and the expectation value of the energy $\varepsilon$; that is, we shall work in a microcanonical ensemble. For this to be meaningful, we must limit ourselves to states whose energy mean-square deviation is small; in other words, to states localized in energy but not necessarily the energy eigenstates. For purposes of calculations we use a well-known family of coherent states\cite{RADCLIFFE, ARECCHI} such that the previous requirement is satisfied. We point out here that the particular state with all particles initially  in one of the wells, analyzed in most of the studies with full quantum solutions\cite{Holthaus,Paredes1,Paredes2},  belongs to the family of such coherent states. 

The first important conclusion is that, for all the above mentioned initial states and for all values of the strength of the particle interactions, the system reaches a state that can be called ``statistically stationary". That is, the distributions of the {\it time-dependent} expectation values $< \hat N_1 >$ and $< \hat J_x \hat J_x >$ are peaked around a value dependent only on $N$ and $\varepsilon$, with a width that becomes relatively smaller as $N$ is increased; in other words, the distribution of those values is sharply peaked. Secondly, we find that the {\it mean} values of those distributions follow the same function as the expectation values of  the operators $\hat N_1$ and $\hat J_x \hat J_x$ in the energy eigenstates, truly stationary states. 

For small values of the interaction strength, typically, the values of  $< \hat N_1 >$ and $< \hat J_x \hat J_x >$  oscillate following the coherent oscillations predicted by the mean-field calculation for a brief period of time, then ``decay" to a stationary value for a long period of time (e.g. $< N_1 >\approx N/2$) followed by revivals. As the number of particles is increased, the time spent in the revival regions and the time spent in the stationary region both increase. However, the ratio of the latter to the former also increases, such that most of the time is spent in the stationary region. Could this process be called {\it decoherence}? It certainly so if experimentally one is able to measure few (i.e. one- and two-) body properties only. After all, this is what statistical physics asserts: any {\it isolated} system left undisturbed for a  long time relaxes to a stationary (equilibrium) state. For larger values of the interaction not only the decay to stationary states occurs, but also the transition to self-trapped states appears. The main result is that there exists a critical value of the interaction,  such that if the interaction is greater than this value, the self-trapping transition occurs as a function of the energy of the system. We recall that since the one-particle spectrum is finite (i.e. just two states), the $N$-body set of allowed energy values is bounded from above and below. 

This article is organized as follows. We briefly present the Hamiltonian of the system and the quantities to be analyzed. We then show results that display the existence of the statistically stationary states and we discuss the transition from delocalized to the self-trapped states. We conclude with a discussion of our results and additional aspects of the phenomenology of this system, such as its characterization in different ensembles. 

\section{The model, observables and eigenstates properties}

We model the interacting quantum fluid confined in the double well through the Bose-Hubbard Hamiltonian considering the two-mode approximation\cite{Milburn}

\begin{equation}
\label{heff}
\hat{\mathcal{H}}=\frac{\hbar \Delta}{2}\left(b_1^\dagger b_2 + b_2^\dagger b_1\right)+ U \left(b_1^\dagger b_1^\dagger b_1 b_1+ b_2^\dagger b_2^\dagger b_2 b_2\right),
\end{equation}
where $\Delta$ is the tunneling frequency of the two lowest energy modes in a symmetrical two well potential and $U= 4\pi \hbar^2 a/m$ represents the effective particle-particle interaction strength  written in terms of the (positive) $s$-wave  scattering length $a$.  We shall specify the different regimes by the adimensional interaction parameter $\Lambda = N U/\hbar \Delta$. The model may also be seen as the collection of $N$ atoms with two internal states in an external static field, and whose pair interaction is long range and only acting if the particles are in the same internal state. We shall return to this version in the last section. We use units with $\hbar = \Delta = m =1$.

As it has been already established, for particular initial states the system described by Hamiltonian (\ref{heff}) exhibits a continuous transition from particle coherent oscillations to a self-trapping regime when the parameter $\Lambda$ is increased from zero to above a critical value $\Lambda_{c}$. This transition may  also occur for a fixed value of $\Lambda$, by varying the initial state, which is equivalent to varying the energy of the system. As a matter of fact, the experiment by Albiez et al.\cite{Oberthaler} using  Bose atoms at very low temperatures, reports the observation of a self-trapped state as a function of the initial population imbalance.

It proves convenient to introduce the one-body operators\cite{Milburn},
\begin{eqnarray}
\hat J_x &=& \frac{1}{2}\left(b_1^\dagger b_2 + b_2^\dagger b_1\right) \nonumber \\
\hat J_y &=& - \frac{i}{2}\left(b_1^\dagger b_2 - b_2^\dagger b_1\right) \nonumber \\
\hat J_z &=& \frac{1}{2}\left(b_1^\dagger b_1 - b_2^\dagger b_2\right)  \label{Js}
\end{eqnarray}
that obey the SU(2) commutation relations, $\left[J_k,J_l\right] = i \varepsilon_{klm} J_m$, and the number of particles $\hat N = b_1^\dagger b_1 + b_2^\dagger b_2$. With these identifications the Heisenberg equations of motion for the $\hat J_k$ operators may be transformed to their mean field counterparts by identifying $\hat J_k \to N K_k $ with $K_k$ a $c$-number function of time. It is believed, that such an approximation is valid for $N \to \infty$, $U \to 0$, but $\Lambda$ finite\cite{Holthaus}. As pointed out in the introduction, this procedure gives information, within the limit considered, on one-body properties only. Nevertheless, this approximation is susceptible of extensions to systems with external time-dependent fields, whose implementation in exact (numerical) quantum calculations may be quite difficult.

By solving for the full $N$-body wavefunction $| \psi_N(t) \rangle$ one can calculate the $N$-body density matrix $\hat \rho^{(N)}(t) = | \psi_N(t) \rangle \langle \psi_N(t) |$, and with this, the reduced one- and two-body density matrices, $\hat \rho^{(1)}$ and $\hat \rho^{(2)}$, whose matrix elements are,
\begin{equation}
\rho^{(1)}_{\alpha \beta} = {\rm Tr}\left(  \hat b_\alpha \> \hat \rho^{(N)} \> \hat b^{\dagger}_\beta \right),
\end{equation}
and
\begin{equation}
\rho^{(2)}_{a b} = {\rm Tr}\left( \hat b_\mu \hat b_\nu \> \hat \rho^{(N)} \> \hat b^{\dagger}_\kappa \hat b^{\dagger}_\eta \right),
\end{equation}
with the Greek subindices taking the values 1 and 2, and $a = (\mu \nu)$ and $b = (\kappa \eta)$, that is, the values $(11)$, $(12)$ and $(22)$. Knowledge of these two operators suffices to determine all one- and two-body properties of the system. Although, we can calculate all the matrix elements of $\hat \rho^{(1)}$ and $\hat \rho^{(2)}$ as a function of time, we find more illustrative to limit ourselves to two properties, one is the number of particles in well 1, namely,
\begin{eqnarray}
\hat N_1 & = & \hat b_1^\dagger \hat b_1 \nonumber \\
&=& \frac{1}{2}\hat N + \hat J_z , \label{N1}
\end{eqnarray}
a one-body operator; and the other is a  ``tunneling correlation",
\begin{equation}
\hat {\cal C} = \hat J_x \hat J_x ,\label{C}
\end{equation}
a two-body operator.

Before studying the time evolution of initial coherent states, we analyze the properties of the eigenstates of the system. First, we numerically solve the eigensystem, $\hat {\cal H}(\Lambda) |\phi_n (\Lambda) \rangle = \varepsilon_n(\Lambda) |\phi_n (\Lambda) \rangle$, where we have indicated that we have a different set for a given $\Lambda$. For $N$ particles there are $N+1$ eigenstates,  the energy of the system is finite and bounded by the lowest $\varepsilon_0(\Lambda)$ and largest $\epsilon_N(\Lambda)$ eigenenergies. We  exhibit all our results, always, for given values of those parameters and as a function of the expectation value of the energy in the given state. In Figs. 1 and 2 we show the values of $< \hat N_1(\Lambda)>_n = \langle \phi_n(\Lambda) | \hat N_1 | \phi_n(\Lambda) \rangle$ and of $< \hat {\cal C}(\Lambda) >_n = \langle \phi_n(\Lambda) | \hat {\cal C} | \phi_n(\Lambda) \rangle$ as a function of the energy $\langle \phi_n (\Lambda)| \hat {\cal H}(\Lambda) | \phi_n(\Lambda) \rangle$, which in this case equals $\varepsilon_n(\Lambda)$.

\begin{figure}
\begin{center}
 \includegraphics[width=1.0\textwidth]{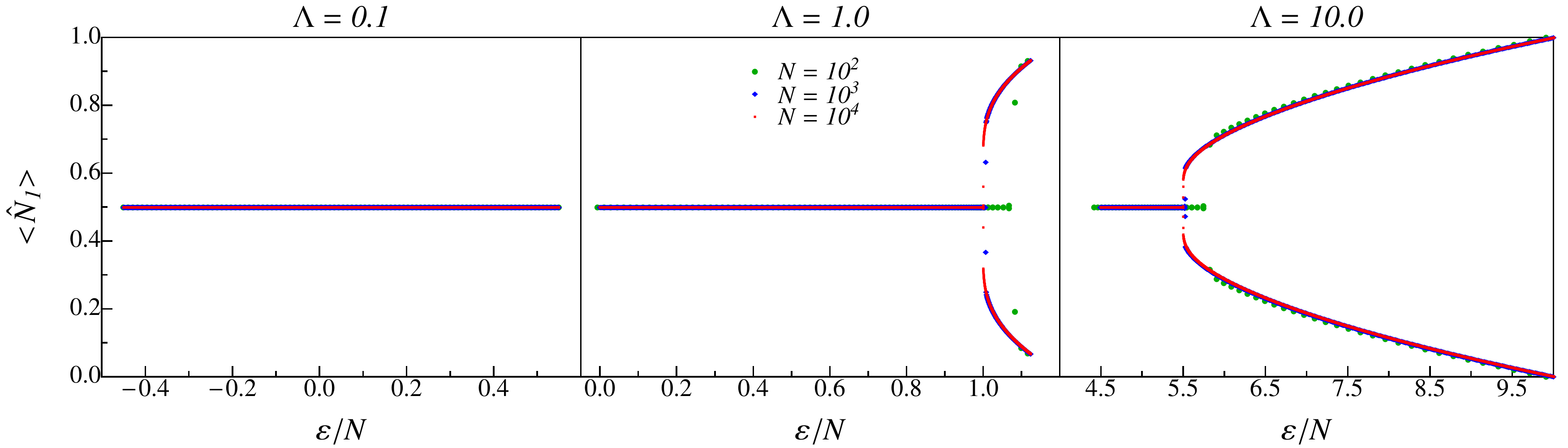}
\end{center}
\caption{(Color online)
Eigenstate expectation value of the particle population in well 1 $\langle N_1 \rangle/N$ as a function of the energy eigenvalues $\varepsilon_n$ for $\Lambda=0.1$, $\Lambda=1.0$ and $\Lambda=10$. Green, blue and red dots,  correspond to $N=$ 100, 1000 and 10000 particles, respectively.}
\label{fig1}
\end{figure}

\begin{figure}
\begin{center}
 \includegraphics[width=1.0\textwidth]{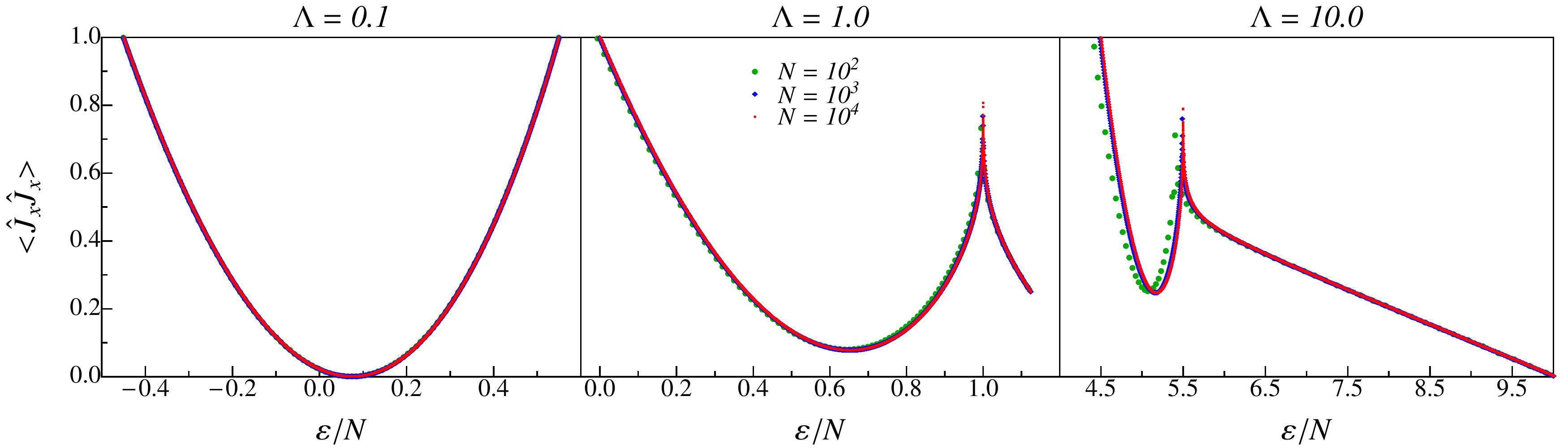}
\end{center}
\caption{(Color online)
Eigenstate expectation value of the two particle correlation ${\cal C}/N^2 = \langle \hat J_x \hat J_x \rangle/N^2$, as a function of the energy eigenvalues $\varepsilon_n$ for $\Lambda=0.1$, $\Lambda=1.0$ and $\Lambda=10$. Green, blue and red dots,  correspond to $N=$ 100, 1000 and 10000 particles, respectively.}
\label{fig2}
\end{figure}

\begin{figure}
\begin{center}
 \includegraphics[width=.7\textwidth]{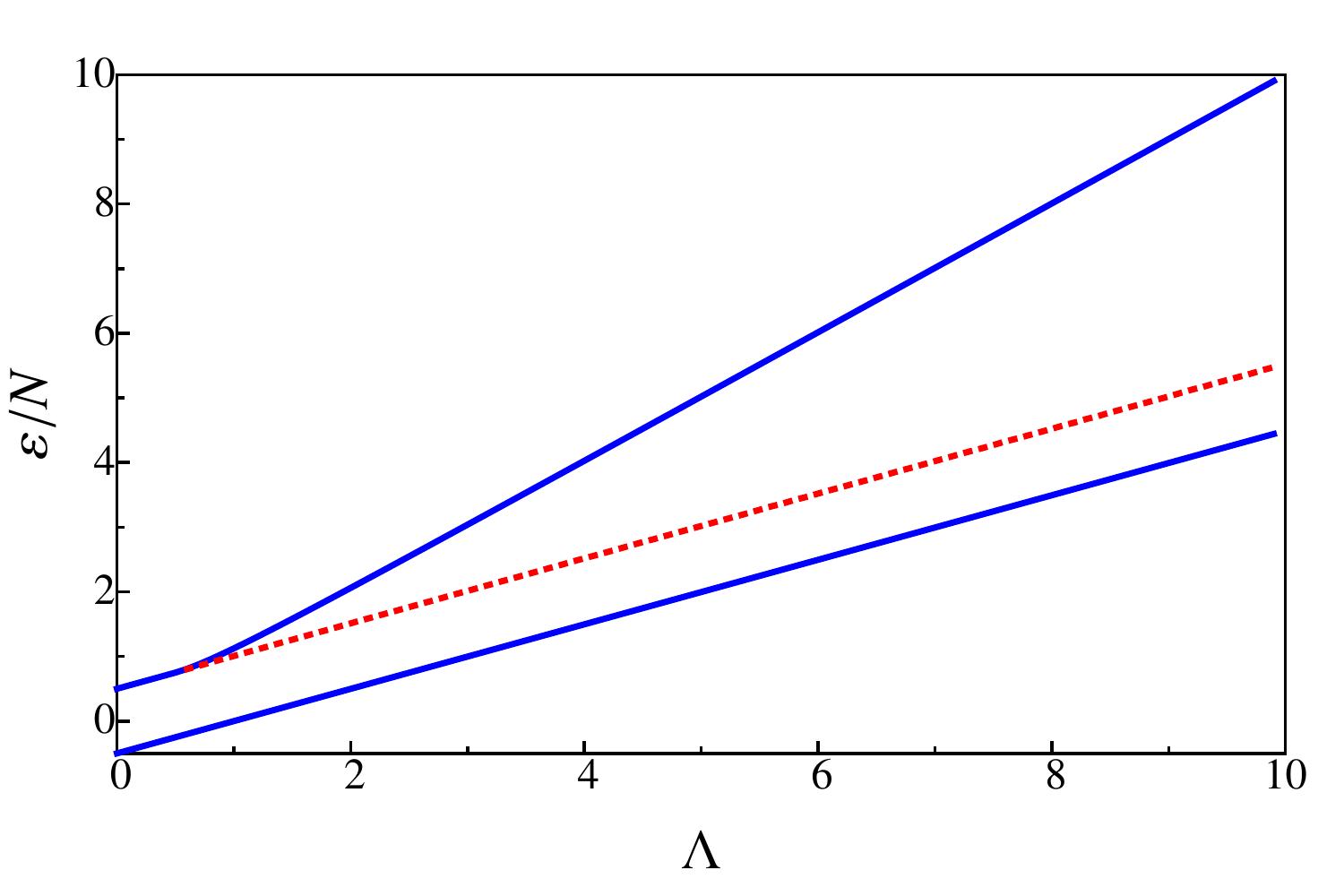}
\end{center}
\caption{Phase diagram $\varepsilon/N$ vs $\Lambda$ for $N = 1000$ particles.  The allowed states are those within the blue lines, $\varepsilon_0(\Lambda)$ and $\varepsilon_N(\Lambda)$. The red line is the transition from delocalized to self-trapped states, $\varepsilon_c(\Lambda)$ for $\Lambda > \Lambda_c \approx 0.6$. The states with energy greater than $\varepsilon_c(\Lambda)$ are self-trapped. }
\label{fig3}
\end{figure}

>From Figs. \ref{fig1} and  \ref{fig2} one can appreciate the intrinsic nature of the stationary eigenstates as $\Lambda$ is varied. For $\Lambda=0.1$ the system does not have any transition, while for $\Lambda=1.0$ and $\Lambda=10.0$, the appeareance of the transition to the self-trapped state is evident. We note that this is a transition as a function of the energy for given $\Lambda$, that is, there is an energy $\varepsilon_c$ above which the system self-traps. In Fig. \ref{fig1} the transition is seen as a ``symmetry breaking" of the eigenstates since the population becomes unbalanced, while in Fig. \ref{fig2} the two-body correlation shows a peak distinguishing between ``delocalized" and self-trapped states. For $N = 1000$ and $N = 10000$ particles the result is essentially the same. As described in previous studies\cite{Paredes1,Paredes2}, in general, when tunneling and interaction energies become comparable $N U \approx \hbar \Delta$, the system is prone to suffer the transition. Here, we find that the transition is continuous, and that occurs at a ``critical" value $\Lambda_{c} \approx 0.6$ for $N = 1000$ particles. All this is best summarized in the phase diagram shown in Fig. \ref{fig3}. As we shall show in the next section, for $\Lambda \ge \Lambda_c$ there exists a ``critical" energy $\varepsilon_c(\Lambda)$ such that, if the energy of the system is above it, $\varepsilon \ge \varepsilon_c$, the system is in the self-trap region, while if its energy is below the transition the system will decay to a delocalized state, regardless of its initial population imbalance. We have dubbed ``delocalized" states as opposite to self-trapped. In the mean-field approach these states are called ``coherent oscillations", however, as described in the next section, the full quantum solution shows a ``decay" to stationary states with half of the population in each well, thus justifying its naming.

\section{Statistically stationary states}

As already pointed out in the literature\cite{Milburn,Holthaus,Paredes1,Paredes2}, the dynamical study of the interacting Bose system with Hamiltonian (\ref{heff}) reveals, in addition to the opposite coherent and self-trapped states, stationary states with recurrent revivals which mean-field treatments are unable to capture. The existence of these stationary states is due, both, to the pair-particle interaction energy and to the fact that we are looking at expectation values of few-body operators. The revivals are consequence of the finite number of eigenstates comprising any conceivable state in this system. To exhibit this phenomenolgy we consider the following family of coherent states\cite{RADCLIFFE, ARECCHI},

\begin{equation}
\label{coher}
|\theta,\phi \rangle = \sum_{n_1=0}^{N} \left(
\begin{array}{c}
 N \\ n_1
 \end{array}\right)^{1/2} \sin^{N-n_1} ({\theta}/{2}) 
\> \cos^{n_1} ({\theta}/{2}) \> e^{-i(N-n_1)\phi}| n_1,N-n_1 \rangle
\end{equation}
where $n_1$ and $N-n_1$ are the number of particles in wells 1 and 2, and $| n_1,N-n_1 \rangle$ are the ($N+1$) Fock states. The coherent states are typically sharply localized both in the Fock states, as well as in the eigenenergy basis. 

For given values of the number of particles $N$, we characterize the coherent states by their energy expectation value, $\epsilon(\theta,\phi) = \langle \theta, \phi | \> {\cal H} \>  | \theta, \phi  \rangle$. Clearly, one can have the same value of $\epsilon(\theta,\phi)$ for different values of $\theta$ and $\phi$. Thus, to simplify the analysis, we choose $\phi \equiv 0$ and vary $\theta$. As mentioned in the Introduction, the state with the $N$ particles in well 1 is given by $\theta = \pi$ and $\phi = 0$. Taken any of these states as the initial one for the system, we evolve it numerically, $| \theta, \phi  ; t \rangle = \exp(-i {\cal H}t/\hbar) | \theta, \phi  \rangle$ and calculate the expectation values of $\hat N_1$ and $\hat {\cal C}$,
\begin{equation}
N_1(t) = \langle \theta, \phi  ; t | \> \hat N_1 \> | \theta, \phi  ; t \rangle \label{N1t}
\end{equation}
and
\begin{equation}
{\cal C}(t) = \langle \theta, \phi  ; t | \> \hat J_x \hat J_x \> | \theta, \phi  ; t \rangle .\label{Ct}
\end{equation}

\begin{figure}
\begin{center}
 \includegraphics[width=1.0\textwidth]{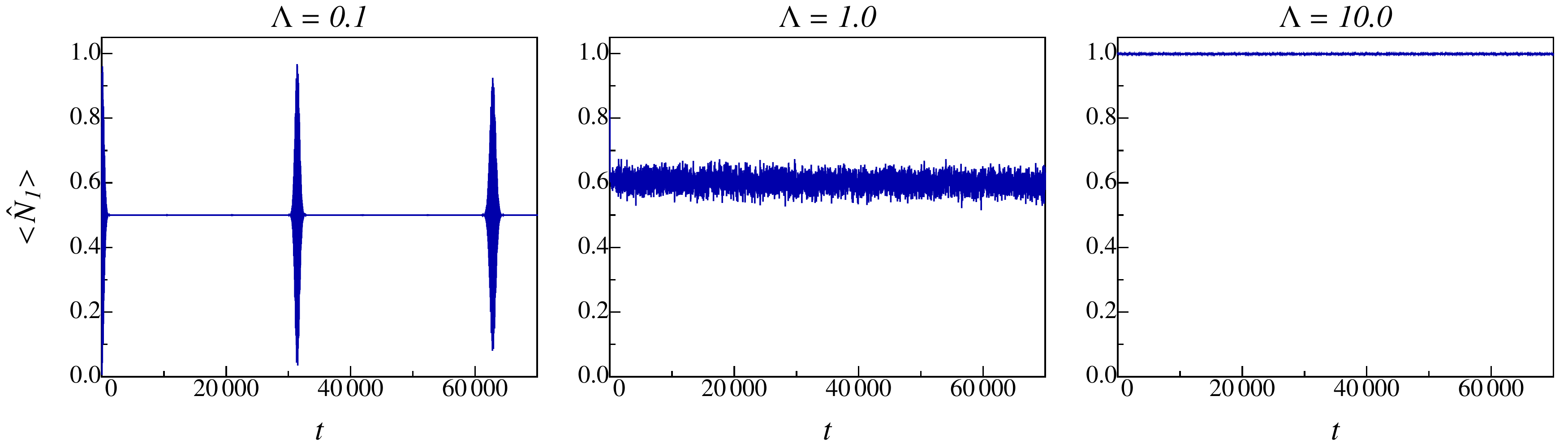}
\end{center}
\caption{Time evolution of the expectation value of the number of particles in well 1,  $N_1(t)/N$, in the coherent state $\theta = \pi$ and $\phi = 0$, for $N = 1000$ and  $\Lambda=$ 0.1, 1.0 and 10.0.}
\label{fig4}
\end{figure}

\begin{figure}
\begin{center}
 \includegraphics[width=1.0\textwidth]{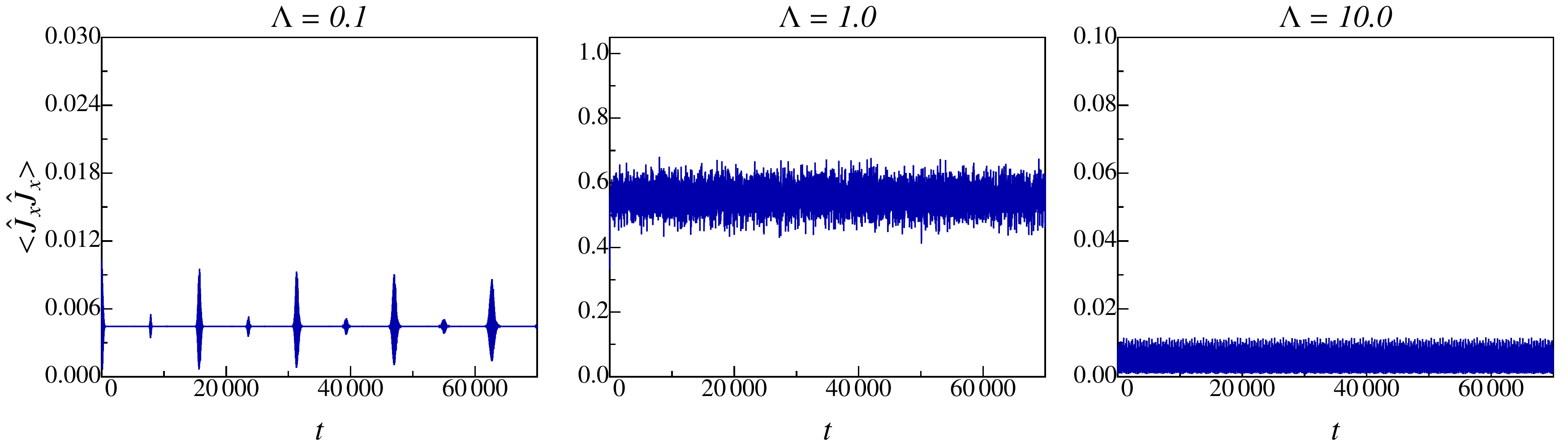}
\end{center}
\caption{Time evolution of the expectation value of the tunneling correlation,  ${\cal C}(t)/N^2$, in the coherent state $\theta = \pi$ and $\phi = 0$, for $N = 1000$ and  $\Lambda=$ 0.1, 1.0 and 10.0. Note the scale in $y$-axis for cases $\Lambda=$ 0.1 and 10.0.}
\label{fig5}
\end{figure}

Figs. \ref{fig4} and \ref{fig5} show the time evolution of the expectation values $N_1(t)$ and ${\cal C}(t)$ in the coherent state $\theta = \pi$ and $\phi = 0$, for $N_1(0) = N$. Figs. \ref{fig3}a and \ref{fig4}a are for $\Lambda = 0.1$. The system shows coherent oscillations initially, then it falls into a stationary value (with $N_1(t)/N = 0.5$) to have revivals at almost periodic intervals. We have numerically verified that the time spent in the stationary regions becomes longer than the time spent in the coherent-oscillation sections, as the number of particles increases. Figs. \ref{fig3}b and \ref{fig4}b correspond to $\Lambda = 1.0$, which are within the crossover transition to the self-trapped states. There are no longer clear coherent oscillations and the mean value of $N_1/N$ is already above 0.5. In Figs. \ref{fig3}c and \ref{fig4}c, corresponding to $\Lambda = 10.0$, the system is well into the self-trap region and the values of $N_1/N$ and ${\cal C}/N^2$ are essentially constant. Below, we explore the whole energy region for the same values of $\Lambda$. The purpose of these figures is to exemplify two important conclusions of our analysis. First,  based on the facts that macroscopically and experimentally we are very much limited to access physical properties of few bodies only, such as energies, temperatures and correlations functions, then, if we can make measurements of, say, $N_1$ and ${\cal C}$ at different times (starting each realization in the same state), one necessarily arrives to the conclusion that the system is, within fluctuations, around a mean stationary value. We shall call these state {\it statistically} stationary states to distinguish them from the true stationary energy eigenstates. This is clearly in agreement with the general principles of statistical physics. We have verified that the behavior shown here for the coherent state  $\theta = \pi$ and $\phi = 0$, is typical for any value of $\theta$ and $\phi$.
The second aspect we want to allude is the  transition from coherent oscillations or delocalized  to self-trapped states as the pair-particle interaction is increased. This point, however, is much better understood with the analysis we provide below.

Once we have shown that the system reaches statistically stationary states, we can describe its properties by the given values of the {\it mean} of $N_1(t)$ and ${\cal C}(t)$, as well as their {\it standard deviation}, for a whole family of coherent states. This family is built in the following manner. For given values of $N$ and $\Lambda$, and taking $\phi = 0$, we find the values of $\theta$ such that the expectation values of the energy in the coherent states span the whole interval $\varepsilon_0(\Lambda) \le \epsilon(\theta,\phi) \le \varepsilon_N(\Lambda)$. Clearly, the value of $\theta$ for a given energy is not unique. We shall see that, depending on this value, the coherent state can break the symmetry. That is, it localizes either near the well 1 or the well 2, once the condition for self-trapping is satisfied, i.e. for $\Lambda \ge \Lambda_c$.

\begin{figure}
\begin{center}
 \includegraphics[width=1.0\textwidth]{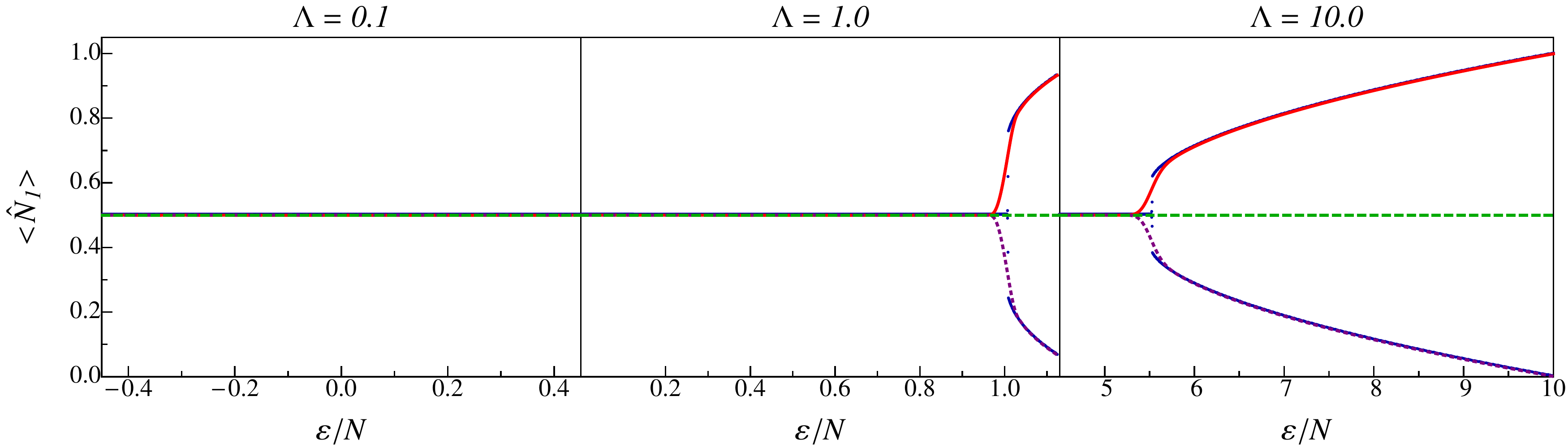}
\end{center}
\caption{(Color online)
Number of particles in well 1,  ${N_1}/N$ as a function of  energy $\epsilon$ of the fluid, for $N = 1000$. The frames correspond to $\Lambda=$ 0.1, 1.0 and 10.0. Blue solid line corresponds to expectation values in energy eigenstates (same as in Fig. \ref{fig1}). Red line corresponds to the time average (mean) of the expectation value in the set of coherent states, Eq.(\ref{coher}) for $\theta(\epsilon)$ such that it localizes in well 1; purple dotted line corresponds to $\pi - \theta(\epsilon)$ that localizes in well 2. Green dotted line corresponds to thermal average values, see Eq.(\ref{terave}).}
\label{fig6}
\end{figure}

\begin{figure}
\begin{center}
 \includegraphics[width=1.0\textwidth]{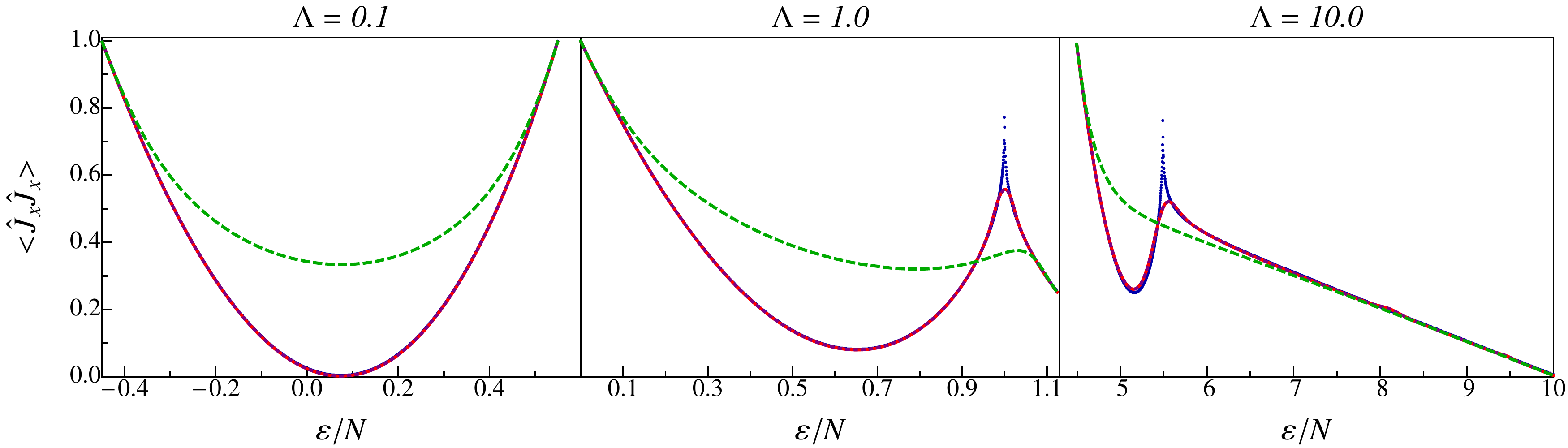}
\end{center}
\caption{(Color online)
Tunneling correlation,  ${\cal C}/N^2$ as a function of  energy $\epsilon$ of the fluid, for $N = 100$. The frames correspond to $\Lambda=$ 0.1, 1.0 and 10.0. Blue solid line corresponds to expectation values in energy eigenstates (same as in Fig. \ref{fig1}). Red line corresponds to the time average (mean) of the expectation value in the set of coherent states, Eq.(\ref{coher}) for $\theta(\epsilon)$ such that it localizes in well 1; purple dotted line corresponds to $\pi - \theta(\epsilon)$ that localizes in well 2. Green dotted line corresponds to thermal average values, see Eq.(\ref{terave}).}
\label{fig7}
\end{figure}

\begin{figure}
\begin{center}
 \includegraphics[width=1.0\textwidth]{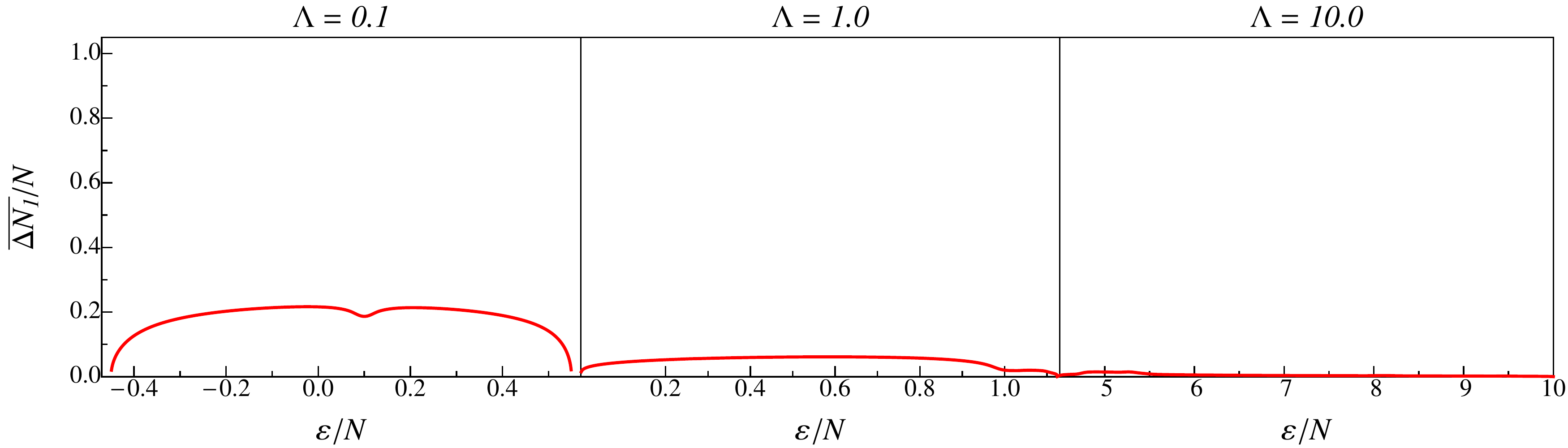}
\end{center}
\caption{Standard deviation (fluctuation) of the expectation value the number of particles in well 1, $\overline{\Delta N_1}/N$ for the set of coherent states, Eq.(\ref{coher}), as a function of energy. $N = 1000$ and  $\Lambda=$ 0.1, 1.0 and 10.0.}
\label{fig8}
\end{figure}

\begin{figure}
\begin{center}
 \includegraphics[width=1.0\textwidth]{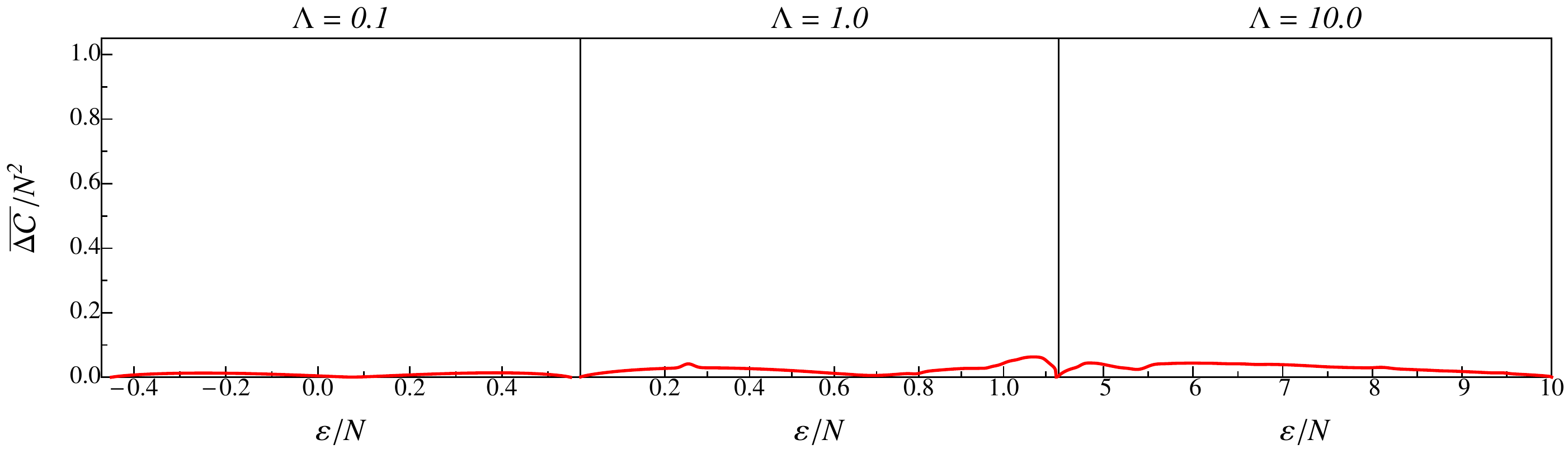}
\end{center}
\caption{Standard deviation (fluctuation) of the expectation value of the tunneling correlation,  $\overline{\Delta {\cal C}}/N^2$, for the set of coherent states, Eq.(\ref{coher}), as a function of energy. $N = 1000$ and  $\Lambda=$ 0.1, 1.0 and 10.0.}
\label{fig9}
\end{figure}

Figs. \ref{fig6} and \ref{fig7} show the mean values of $N_1(t)$ and ${\cal C}(t)$ for $N = 1000$ and for $\Lambda =$ 0.1, 1.0 and 10.0 as a function of the expectation value of the energy of the state. In the same graph we have included the eigenstates expectation values of Figs. \ref{fig1} and \ref{fig2}. In general, we see that the mean values from the coherent states agree fairly well with the eigenstate values. The main deviations are near the critical transition values of the energy; from calculations with $N = 100$ up to $N = 10000$, we can assert that these deviations should become smaller as the number of particles is increased, and should become equal in the thermodynamic limit. Hence, we conclude that the stationary states follow the same behavior as the eigenenergy states and, therefore, that their macroscopic behavior is described by the phase diagram shown in Fig. \ref{fig3}; that is, there exists a critical value $\Lambda_c$, such that, for values above such  a threshold the system exhibits a self-trapping transition as a function of the energy (or as a function of the initial state). Additionally, the states may be well characterized by both one- and two-body properties; that is, while the one-body properties remain constant below the self-trapping transition, two-body variables serve to discriminate among different states. Both quantities clearly signal the transition point.  Figures \ref{fig8} and \ref{fig9} show the standard deviation of the values of $N_1(t)$ and ${\cal C}(t)$; their smallness justify the characterization of the states as stationary.

The transition from delocalized to self-trapped states exhibits a symmetry-breaking phenomenon. That is, the Hamiltonian is symmetric under the exchange of the wells, or internal states, 1 and 2. However, the stationary states ``choose" one of wells to become localized. To be precise, one can show that for $\Lambda \ge \Lambda_c$, if $\phi = 0$ and $\theta$ yields a state localized in well 1, then $\theta^\prime = \pi - \theta$ and $\phi =0$, localizes in well 2. This is shown in Fig. \ref{fig6} for $\Lambda = 10.0$ and $N = 100$, red line signals localization in well 1 and purple line in well 2.

\section{Discussion and final remarks} 

We have analyzed one- and two-body properties of the full quantum solution of the two-mode Bose-Hubbard fluid. We have discussed the transition from delocalized to self-trapped states, which occurs as the energy of the system is increased, provided that the pair-interaction strength is above a critical or threshold value. The full quantum solution for a large number of atoms differs from the mean-field approximation, since the latter predicts coherent Josephson-like oscillations while the exact solution shows that those oscillations decay to stationary like states in which the system spends most of its time. Since, measuring or having access to the $N$-body wave function appears as an impractical task in real systems, most of our understanding of macroscopic systems is based on knowledge of properties of few bodies. In this context, we argue that the decay, or relaxation, to a stationary state can be considered as decoherence, even if no interaction with an external environment is included. Such a decoherence is a consequence of the interatomic interactions. In a way, this is how statistical physics has been developed. That is, one of its basic tenants is the assumption that any interacting, macroscopic closed or isolated system, classical or quantum, relaxes to an stationary state if left undisturbed\cite{Landau}.

Taking the statistical description of the stationary states as an approach to characterize them, we have studied the number of particles in well 1 ${\hat N_1}$ and the ``tunneling correlation" $\hat {\cal C}$, see Eqs.(\ref{N1}) and (\ref{C}), as examples of one- and two-body properties. The phase diagram of Fig. \ref{fig3} summarizes our results. For a given values of the number of particles $N$ and the interaction strength $\Lambda$, we find a transition from delocalized to self-trapped states as a function of the energy $\varepsilon$ of the fluid, if $\Lambda \ge \Lambda_c$, a critical interaction value. One finds that ${N_1}$ remains constant, ${N_1} = 0.5 N$, up to a critical energy value $\varepsilon_c(\Lambda)$, where it changes continuously up to either a value ${N_1} = 0$ or ${N_1} = N$, as the energy is further increased. Likewise, the transition is registered by ${\cal C}$ with a cusp at the transition energy, clearly dividing two different types of macroscopic states. This transition appears to us as the reminiscent of a continuous phase transition with an spontaneous symmetry breaking mechanism\cite{Ma}. The behavior of ${N_1}$ seems analogous to the order parameter of the transition. This is further supported by the suggestion, based on our calculations, that while the change of ${N_1}$ is continuous, its derivative appears discontinuous at the transition, and we expect that should become so in the thermodynamic limit. The symmetry breaking is also clear: while the Hamiltonian is invariant under the interchange ``1" and ``2" of the wells, stationary states with the same energy are not; a set of states localizes in well 1 and another in well 2. This is quite similar to the spontaneous magnetization of a ferromagnetic material\cite{Landau}.

To conclude, and because of the statistical approach we are putting forward, we would like to briefly discuss the role of the temperature in this problem.  First, we recall that an isolated system is best described by a microcanonical ensemble, and in fact, this is how we have described the system. Namely, its macroscopic state is characterized by the number of particles $N$ and the energy $\varepsilon$. Depending on the values of these variables, and if the system is in equilibrium, it should have a well defined temperature $T = T(N, \varepsilon)$. If the ensembles are equivalent, one should obtain the same results for the equilibrium states using the canonical ensemble with $N$ and $T$ given. Due to the simplicity of the system at hand, one can very easily calculate the average value of any observable operator $\hat A$  in the canonical ensemble,
\begin{equation}
\langle \hat A \rangle = \frac{1}{Z} \sum_{n = 0}^N \langle \phi_n | \hat A | \phi_n \rangle \> e^{- \epsilon_n/kT} ,\label{terave}
\end{equation}
where $k$ is Boltzmann constant and the partition function is
\begin{equation}
Z =  \sum_{n = 0}^N \>  e^{- \epsilon_n/kT} .
\end{equation}
With this procedure, and assuming that the temperature is a single valued function of $N$ and $\epsilon$, we can assign a temperature to a stationary state with a given energy $\epsilon$ by solving
the following equation for $T$,
\begin{equation}
\epsilon = \frac{1}{Z} \sum_{n = 0}^N \epsilon_n \> e^{- \epsilon_n/kT} . \label{eter}
\end{equation}
In this way, we can express the thermal averages of $\hat N_1$ and $\hat {\cal C}$ by means of  Eq.(\ref{terave}) as a function of $\epsilon$ and $N$, using the temperature $T(\epsilon,N)$ as described above. The green lines in Figs. \ref{fig6} and \ref{fig7} show one of these examples. The results of the correlation tunneling ${\cal C}$ seem to indicate that there is no agreement between the ``microcanonical" and the ``canonical" descriptions. In Fig. \ref{fig6} one finds that the thermal average of $\hat N_1$ is always $N/2$, whether above or below the transition. This is not a discrepancy since the thermal canonical average cannot show the spontaneous symmetry breaking; it is analogous to the fact that in the Ising model in 2D the average magnetization is always zero:  one needs a small external magnetic field to break the symmetry and then make it to vanish\cite{Huang}. The somewhat perplexing problem is in the behavior of the average values of the two-body tunneling correlation $\hat {\cal C}$, since the peak is not shown in the thermal averages. We do not have an explanation for this discrepancy. It may either be that our calculation based on the eigenstates and the coherent states do not correspond to a microcanonical ensemble, or that for this system there is simply no equivalence between ensembles. The latter option cannot at all be disqualified. In a way this system is peculiar since it represents a collection of atoms interacting with the same strenght with all the other atoms, provided they are in the same internal state. In other words, the fact of considering the two-mode approximation only has  given rise to an effective Hamiltonian in which the interactions among the particles appear as if they were long range, whenever the particles have the same internal state, while the full Hamiltonian represents particles with short range interactions. Moreover, it can also be shown, by studying Eq.(\ref{eter}), that the energy as a function of temperature and number of particles is not fully extensive, namely, it is not of the form $\epsilon(T,N) = N \> e(T)$ with $e(T)$ a function of the energy only, but rather, $\epsilon(T,N) = N e(T/N)$. The elucidation of these discrepancies deserves a study on its own.

{\bf Acknowledgments}. This work was partially supported by grant IN114308 DGAPA (UNAM). Computational facilities by DGSCA-UNAM are also acknowledged.


\begin{thebibliography}{99}
\bibitem{eigenvalues} Based on the calculation of the eigenvalues we found that this condition is satisfied.
\end{thebibliography}
\end{document}